\documentclass[12pt, a4]{iopart}

\usepackage{iopams}  
\usepackage[dvips]{graphicx,color}

\usepackage{endnotes}
\let\footnote=\endnote

\begin{document}

\title[A. Miyazaki et al]{Determination of the BCS material parameters of the HIE-ISOLDE superconducting resonator}

\author{A. Miyazaki$^{1,2}$ and W. Venturini Delsolaro$^{1}$}

\address{$^1$CERN, Switzerland}
\address{$^2$University of Manchester, UK}
\ead{Akira.Miyazaki@cern.ch}
\vspace{10pt}
\begin{indented}
\item[]May 2018
\end{indented}

\begin{abstract}
Superconducting material parameters of the Nb film coating on the Quarter-Wave Resonator (QWR) for the HIE-ISOLDE project were studied by fitting experimental results with the Mattis-Bardeen theory.
We pointed out a strong correlation among fitted estimators of material parameters in the BCS theory, 
and proposed a procedure to remove the correlation by simultaneously fitting the surface resistance and effective penetration depth.
Unlike previous studies, no literature values were assumed in the fitting.
As surface resistance and penetration depth had a similar dependence on coherence length and mean free path,
the correlation between these two parameters could not be eliminated by this fitting.
The upper critical field measured by SQUID magnetometry showed complementary constraint to the RF result, and this allowed all the material parameters to be determined.
\end{abstract}

%
\vspace{2pc}
\ioptwocol

\section{Introduction}
The HIE-ISOLDE Linac~\cite{hie_isolde} is an upgrade project for post-acceleration of heavy ions produced in the ISOLDE facility at CERN~\cite{ISOLDE}.
The Linac is composed of four cryomodules, each containing five superconducting Quarter-Wave Resonators (QWR)~\cite{QWR} and one focusing solenoid.
The QWR is made of a thin Nb film a few microns thick, deposited on a Cu substrate.
A DC-bias sputtering method has been used to coat the cavity, and gave rise to a fine crystal structure with a lot of grain boundaries and dislocations in the film~\cite{alban:sputtering}.
Therefore, determination of material parameters without relying on the literature values for clean bulk Nb is of great importance.

\section{BCS fitting}
The surface impedance of a superconductor at low Radio Frequency (RF) fields was initially obtained from linear response theory.
Mattis and Bardeen (MB) were the first to apply the first order perturbation derived in the original paper by Bardeen Cooper and Schrieffer (BCS)~\cite{BCS} to the RF response including the anomalous skin effect~\cite{MB}.
Abrikosov, Gor'kov and  Khalatnikov also derived the same result in the clean limit by using the formalism of Green functions~\cite{abrikosov}.
This formalism provided a more systematic way to treat impurities considering the Born approximation of the scattering potential caused by homogeneously distributed scattering centers.
Halbritter developed a general algorithm applicable to arbitrary mean free path but RF frequency below half of the superconducting gap~\cite{halbritter1}.
He also implemented its numerical calculation by FORTRAN66~\cite{halbritter2}.
In this study, a C\texttt{++} code has been developed for parallel computing in the Linux cluster at CERN (LXPLUS).
C\texttt{++} was selected in order to achieve both precision and reasonably short computation time.

The numerical code requires the following five input parameters:
\begin{enumerate}
\item BCS coherence length $\xi_0$\footnote{The original code~\cite{halbritter2} used $\xi_{\rm F} = \pi\xi_0/2$ instead. This is a characteristic length of the Gor'kov functions.}
\item London penetration depth $\lambda_{\rm L}$
\item mean free path of normal conducting electrons $l$
\item coupling constant $\Delta_0/k_{\rm B}T_{\rm c}$
\item critical temperature $T_{\rm c}$
\end{enumerate}
and returns two physics quantities as output: \footnote{
In this study, one of the two extreme boundary conditions, diffuse boundary, was selected. 
This corresponds to a random scattering of electrons at the surface, and was considered as more realistic than the other where electrons reflect specularly.
These two boundary conditions were studied by expanding the Boltzmann equation up to the first order of the electric field~\cite{boundary}.
Although the most realistic boundary condition may be between these two extreme cases,
practically, one can select the former,
because the difference between these two boundary conditions is typically less than $10$~\% in the surface resistance.
}
\begin{enumerate}
\item surface resistance $R_{\rm BCS}$ (BCS resistance)
\item effective penetration depth $\lambda_{\rm BCS}$
\end{enumerate}
as functions of two macroscopic variables:
\begin{enumerate}
\item temperature $T$,
\item RF frequency $f$ (appears only in $R_{\rm BCS}$).
\end{enumerate}
If the experimental data are to be explained by the BCS theory, ideally, this code should be able to fit the data and determine the input material parameters.

\subsection{Experimental data}
The cavity was measured 
in a vertical cryostat, and had two ports, one for the fundamental power coupler and the other for the pick-up antenna.
The coupler was mobile and the external coupling was controlled by a stepper motor.
The accelerating field $E_{\rm acc}$ and quality factor $Q_0$ were obtained from the time constant $\tau$ of the field decay, the forward power $P_{\rm f}$, the reflected power $P_{\rm r}$, and the transmitted power through the pick-up port $P_{\rm t}$.
The RF measurement is summarized in \ref{sec:RF_meas}.

The surface resistance $R_{\rm s}$, averaged over the cavity surface, was obtained by the measured $Q_0$
\begin{equation}
R_{\rm s} = \frac{G}{Q_0},
\end{equation}
where the geometrical factor $G$ was evaluated by the commercial codes CST MICROWAVE STUDIO~\cite{CST} and HFSS~\cite{HFSS} and is $30.8\, \Omega$ for the HIE-ISOLDE QWR.
This $R_{\rm s}$ was measured as a function of cavity temperature typically between $2.3$~K and $4.6$~K, which was determined by 
our cryogenic system.
The measurement was done near the critical coupling condition, and $E_{\rm acc}$ was kept constant by controlling $P_{\rm f}$.
Empirically, the measured $R_s(T)$ at a low RF field can be decomposed into two terms
\begin{equation}
\label{eq:two_fulid_Rs_vs_T}
R_s(T) = \frac{A\omega^{n}}{T}\exp{\left(-\frac{\Delta_0}{k_{\rm B}T} \right)} + R_{\rm res},
\end{equation}
where $A$ depends on material parameters, $\omega$ is angular resonance frequency, $1.5<n<2.0$, and $R_{\rm res}$ is a temperature independent component called residual resistance.
The first term in (\ref{eq:two_fulid_Rs_vs_T}) looks similar to the formula derived by the BCS-MB theory approximated by constant Matrix elements~\cite{halbritter3}, 
and thus is usually equated to the BCS resistance.
In this study, this temperature dependent component was fitted by the full BCS-MB theory \footnote{
If there is no reasonable fitting solution, this exponentially temperature dependent term might require extension from conventional linear response theory.
The residual resistance is only empirically temperature-independent in this context, but could weakly depend on temperature and might contaminate the BCS term.
A unified theory including both BCS and weakly temperature-dependent $R_{\rm res}$ was proposed~\cite{Kubo_unify};
however, the result, based on the Usadel equation, is only valid in the dirty limit.}.

The resonance frequency of the phased-locked cavity was typically measured between $7$~K and $9.5$~K during warming up.
For this measurement, the mobile coupler was placed in an over-coupled condition ($P_{\rm r}/P_{\rm f}\sim0.8$ at $7$~K) 
so that the Phase Lock Loop (PLL) worked properly even for the low power transmission present at higher temperature.
Once the Data AcQuisition (DAQ) started, no parameters were controlled so as to reduce the possible systematic errors on the resonance frequency.
The change in effective penetration depth $ \lambda$ can be obtained by the frequency shift $\Delta f$ using Slater's theorem~\cite{lambda_to_f}
\begin{eqnarray}
\Delta\lambda & = & -\frac{G}{\pi\mu_0f^2}\Delta f,
\end{eqnarray}
with,
\begin{eqnarray}
\Delta \lambda & = & \lambda(T) - \lambda(T_0) \\
\Delta f & = & f(T) - f(T_0),
\end{eqnarray}
where $T_0$ is the starting temperature of DAQ.

\begin{figure}[ht]
\includegraphics[width=90mm]{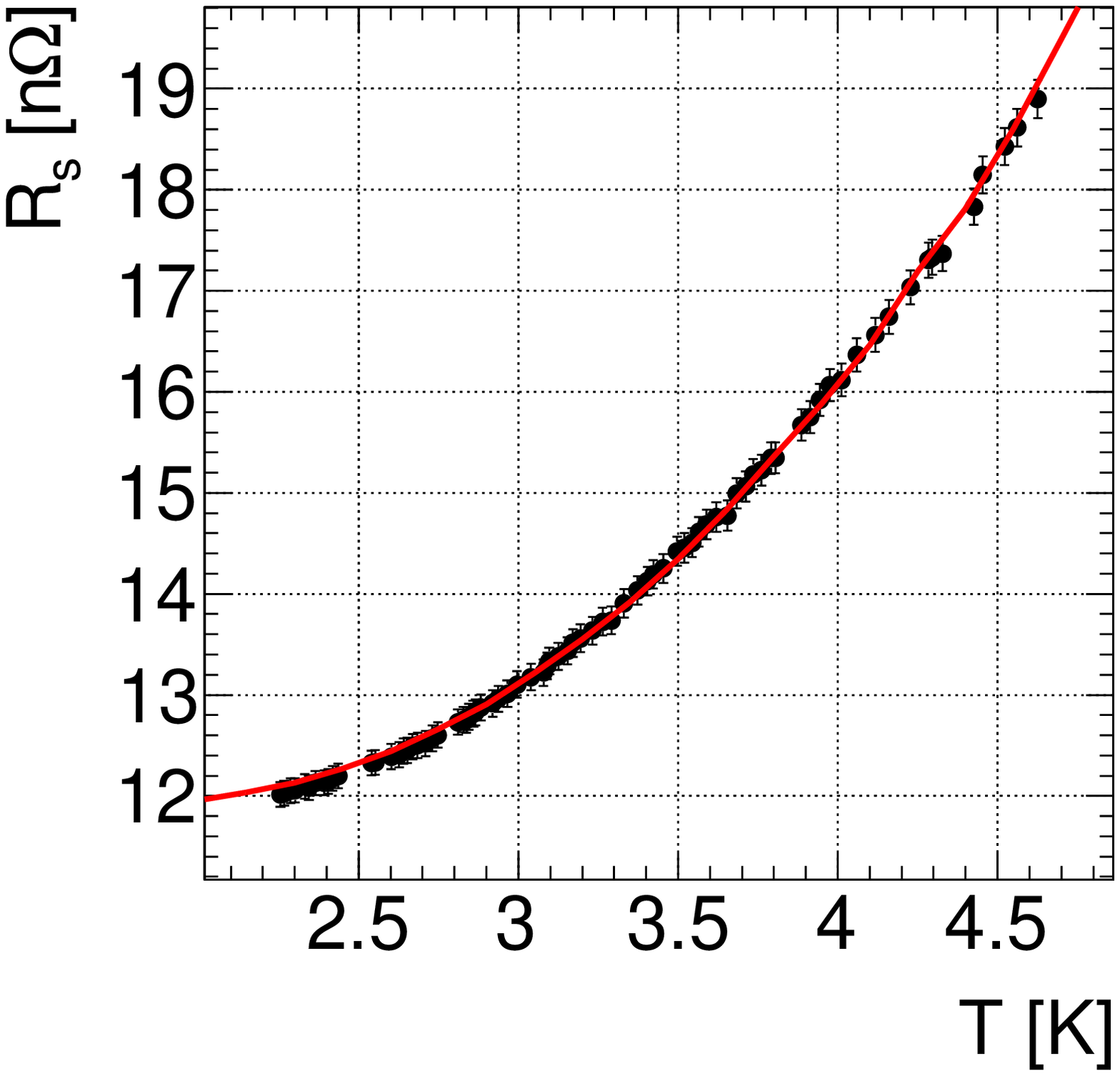}
\caption{
Surface resistance as a function of temperature. The dots show data points and the solid is the line best fit.
\label{fig:Rs_vs_T}}
\includegraphics[width=90mm]{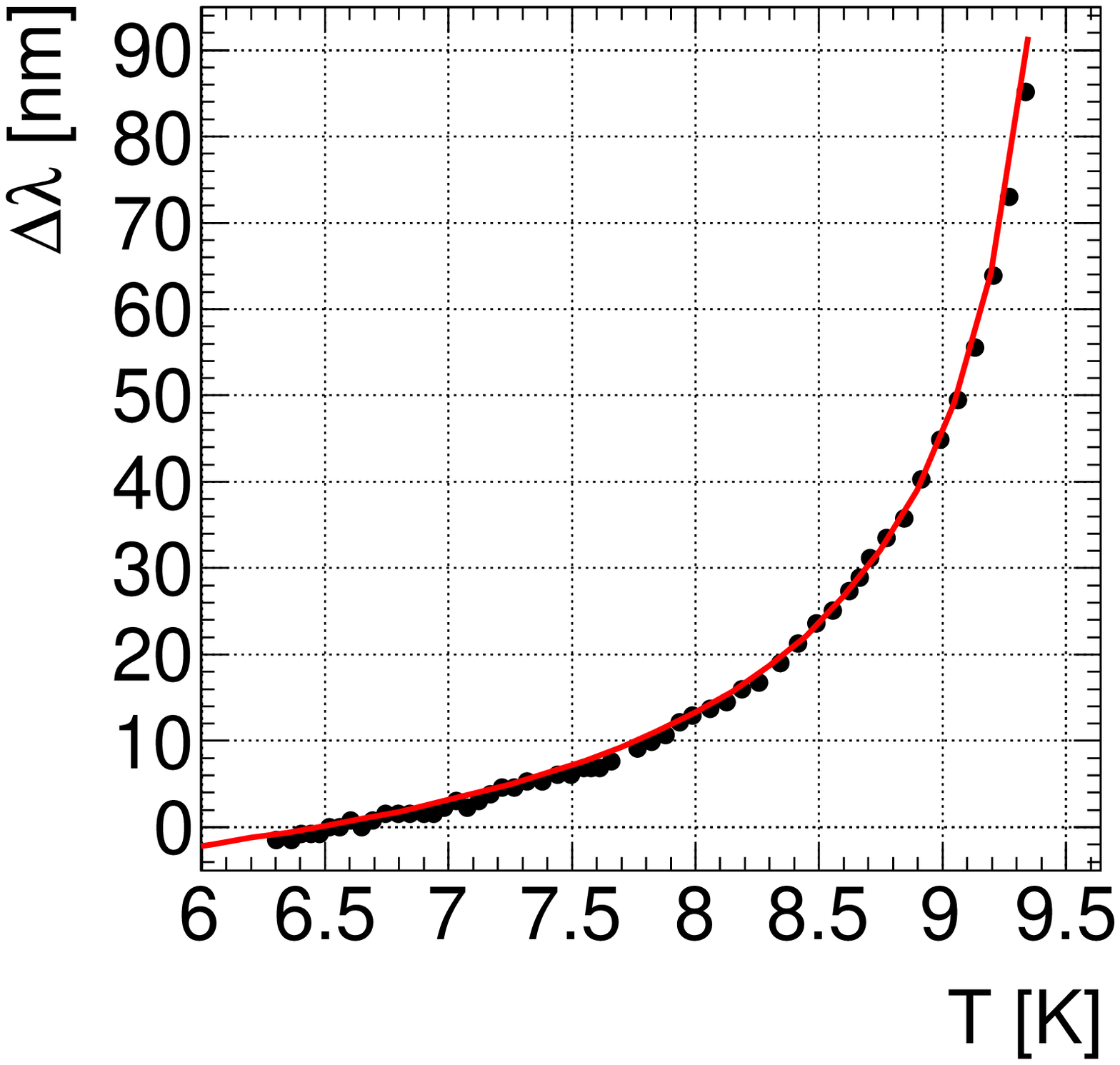}
\caption{
Shift in effective penetration depth as a function of temperature. The dots show data points and the solid line is the best fit.
\label{fig:lambda_vs_T}}
\end{figure}

Figure~\ref{fig:Rs_vs_T} and  \ref{fig:lambda_vs_T} show typical measurement data.
The solid lines shown on the data points are the results of the fitting reported in this study.
Special care was taken in order to eliminate non-BCS-MB phenomena.
First, $E_{\rm acc}$ was kept at a low field (accelerating field of $0.3$~MV/m or peak magnetic field of $3$~mT) to avoid the systematic errors caused by the Q-slope problem~\cite{NbCu}.
Also, the measurement was done after a thermal cycle during which the cavity was cooled as slowly as possible in order to achieve a uniform temperature distribution on the surface~\cite{thermal_cycle}.

\subsection{Correlation among parameters of the BCS impedance}
The fitting procedure was as follows.
First, $R_{\rm res}$ was removed by fitting the empirical formula shown in (\ref{eq:two_fulid_Rs_vs_T}).
The BCS-MB term of the data was then defined as
\begin{equation}
R_{\rm data} = R_{\rm s}-R_{\rm res}
\end{equation}
Second, the critical temperature $T_{\rm c}$ was determined by a Meissner effect measurement using the flux-gate sensors around the cavity.
Third, the starting temperature of the frequency measurement $T_0$ was fixed.
Finally, the temperature dependent surface resistance $R_{\rm data}$ and shift in effective penetration depth $\Delta \lambda$ were fitted
by the numerical calculation of BCS theory to determine remaining four free material parameters $\left(\xi_0, \lambda_L, l, \Delta_0/k_{\rm B}T_{\rm c} \right)$.
The fitting could be done by minimizing $\chi^2$ defined by
\begin{eqnarray}
\chi^2(R_{\rm s}) & = & \sum_{j=0}^{n_{R_{\rm s}}} \left[ \frac{R_{\rm data}(j) - R_{\rm BCS}(T_{\rm j})}{\sigma_{R_{\rm s}}(j)} \right]^2 \\
\chi^2(\lambda) & = & \sum_{j=0}^{n_{\lambda}} \left[ \frac{\Delta\lambda(j) - \Delta\lambda_{\rm BCS}(T_{\rm j})}{\sigma_{\lambda}(j)} \right]^2,
\end{eqnarray}
where $n_{R_{\rm s}}$ and $n_{\lambda}$ are the number of data points of surface resistance and penetration depth measurement, respectively,
and $\sigma_{R_{\rm s}}(j)$ and $\sigma_{\lambda}(j)$ are their associated standard deviations at point $j$ \footnote{
These standard deviations contain systematic uncertainties commonly shared by all the data points, 
as well as fluctuations of the data do not necessarily follow Gaussian distributions.
Therefore, the $\chi^2$ defined here does not necessarily obey the proper $\chi^2$ distribution, and confidence intervals may not be well defined.
However, the minimum of the $\chi^2$ can still reliably determine the best fitting parameters.}.
There are two sets of parameters, one which gives the minimum $\chi^2(R_{\rm s})$ and one which gives the minimum $\chi^2(\lambda)$.
These two results should be consistent and should be averaged, or should converge into one solution after some iterations.

However, the strategy described above did not work well.
The strong correlations which exist between estimators for the fitting parameters $\left(\xi_0, \lambda_L, l, \Delta_0/k_{\rm B}T_{\rm c} \right)$ prevented a non-linear minimizer from finding the solution.
In order to see this effect, a grid search was done using Linux cluster.
A job for a single CPU was coded as
\begin{itemize}
\item $l$ scan: $80$ points between $25$~nm and $185$~nm
\item $\Delta_0/k_{\rm B}T_{\rm c}$ scan: $10$ points between $1.5$ and $2.5$
\item $\chi^2$ calculation for $n_{R_{\rm s}}\sim 80$ and $n_{\lambda}\sim 50$.
\end{itemize}
This single job took several hours by Intel Core Processor i7 (Haswell, no TSX). 
The jobs were distributed to 700 CPUs by
\begin{itemize}
\item $\xi_0$ scan: $35$ points between $10$~nm and $45$~nm
\item $\lambda_{\rm L}$ scan: $20$ points between $20$~nm and $40$~nm
\end{itemize}

\begin{figure}[ht]
\includegraphics[width=80mm]{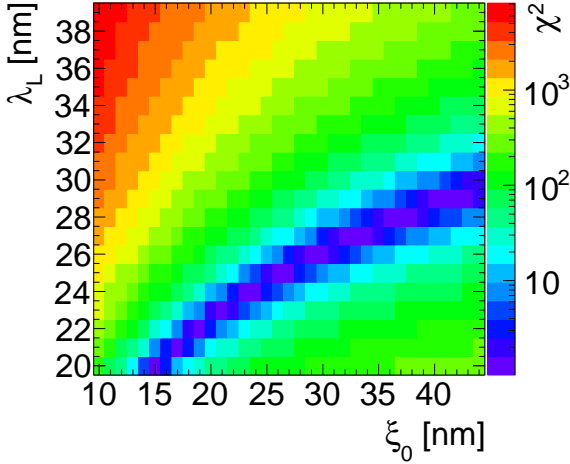}
\caption{
$\chi^2$ distribution of surface resistance normalized by the minimum value. The other parameters are $l=99$~nm and $\Delta_0/k_{\rm B}T_{\rm c}=1.7$.
\label{fig:lambdaL_vs_xi0_Rs}}
\end{figure}

A typical result of the $\chi^2$ distribution for the surface resistance is shown in Fig.~\ref{fig:lambdaL_vs_xi0_Rs}.
Note that this is only a two dimensional cross-section of a four dimensional hyper-surface of the input parameters.
There is a {\it valley} of minimum $\chi^2$ and therefore the standard minimizer calculating partial differentials of $\chi^2$ by the parameters got lost.
This is a demonstration of the strong correlations among estimators of parameters \footnote
{
As a sanity check, a dummy data set produced by the BCS code itself was fitted by the same code. 
The $\chi^2$ converges to zero with the true parameters, but the region of minimum is very narrow and immediately smeared by finite experimental fluctuations.
}.
In this example, the correlation between $\lambda_{\rm L}$ and $\xi_0$ values indicates that a curve of the function $R_{\rm BCS}$ modified by higher $\lambda_{\rm L}$ can be recovered by higher $\xi_0$, and they are thus uncertain.
In other words, it is easily possible that the best fitted parameters from the surface resistance are far from those independently determined by the penetration depth.

\subsection{Simultaneous fitting by $R_{\rm BCS}$ and $\lambda_{\rm BCS}$}
\begin{figure}[ht]
\includegraphics[width=80mm]{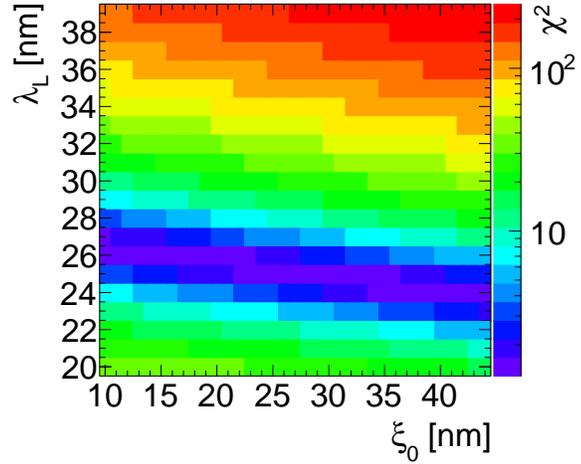}
\caption{
$\chi^2$ distribution of penetration depth normalized by the minimum value. The other parameters are $l=99$~nm and $\Delta_0/k_{\rm B}T_{\rm c}=1.7$.
\label{fig:lambdaL_vs_xi0_lambda}}
\end{figure}

As another group reported~\cite{Cornell_PhD}, combining surface resistance and penetration depth can mitigate the correlations.
Figure~\ref{fig:lambdaL_vs_xi0_lambda} shows the $\chi^2$ distribution obtained by the penetration depth measurement.
As the function of $\lambda_{\rm BCS}$ depends on $(\xi_0, \lambda_{\rm L})$ differently from that of $R_{\rm BCS}$,
the {\it valley} of minimum $\chi^2$ has an intersection with Fig.~\ref{fig:lambdaL_vs_xi0_Rs} near the center of the plot.
Such an intersection may be a robust guess of the true solution in this two dimensional surface.

\begin{figure}[ht]
\includegraphics[width=80mm]{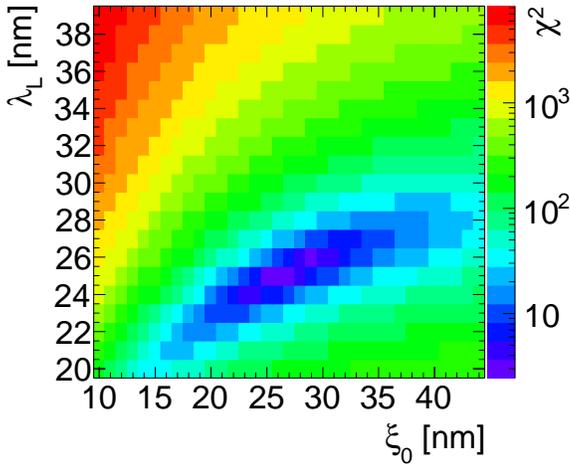}
\caption{
Merged $\chi^2$ distribution of surface resistance and penetration depth. The other parameters are $l=99$~nm and $\Delta_0/k_{\rm B}T_{\rm c}=1.7$.
\label{fig:lambdaL_vs_xi0_combined}}
\end{figure}

In order to obtain the intersection, we considered the sum of the two independent $\chi^2$s\footnote
{
The normalization by the minimum $\chi^2$ enforces the same weight or significance to the two different data.
In this analysis, the statistical and systematic errors in the data are overwhelmed by the uncertainty due to the structure of the parameter correlations.
The $\chi^2$ variation around the global minimum of this modified $\chi^2$ may not give the statistically well defined confidence intervals.
However, this $\chi^2$ still provides a reasonable indicator of the uncertainty of the parameters associated with the correlations.
}
\begin{equation}
\chi^2 \equiv \frac{\chi^2(R_{\rm s})}{\min \left\{ \chi^2(R_{\rm s}) \right\}} + \frac{\chi^2(\lambda)}{ \min \left\{ \chi^2(\lambda) \right\}}.
\end{equation}
The result is shown in Fig.~\ref{fig:lambdaL_vs_xi0_combined}.
The problematic {\it valley} of the $\chi^2$ minimum becomes smaller, where two parameters $(\xi_0, \lambda_{\rm L})$ were more precisely confined.
This simultaneous fit worked well for most of the sets of parameters: $(\xi_0, \lambda_{\rm L})$, $(\lambda_{\rm L}, l)$, $(\xi_0, \Delta_0/k_{\rm B}T_{\rm c})$, $(\lambda_{\rm L}, \Delta_0/k_{\rm B}T_{\rm c})$, and $(l, \Delta_0/k_{\rm B}T_{\rm c})$.

\begin{figure}[ht]
\includegraphics[width=80mm]{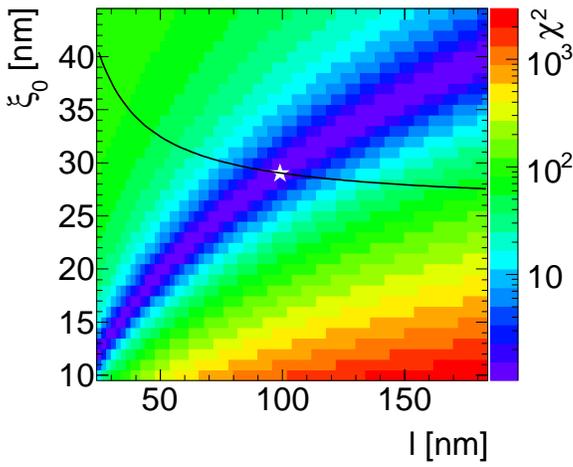}
\caption{
Merged $\chi^2$ distribution of surface resistance and penetration depth. The other parameters are $\lambda_{\rm L}=26$~nm and $\Delta_0/k_{\rm B}T_{\rm c}=1.7$.
The solid line is explained in Sec.~\ref{sec:mag}
\label{fig:xi0_vs_l_combined}}
\end{figure}

However, the correlation of $(\xi_0, l)$ cannot be solved by this method as shown in Fig.~\ref{fig:xi0_vs_l_combined}.
This is because both the surface resistance and the penetration depth depend on $(\xi_0, l)$ similarly.
The well confined result of $(\xi_0, \lambda_{\rm L})$ and others is just one of the cross-sections of this remaining {\it valley}, and not the true solution.

In previous studies on bulk Nb, Nb$_3$Sn and N-doped bulk Nb~\cite{Cornell_PhD}, 
this correlation was eliminated by using literature values for $(\xi_0, \lambda_{\rm L})$.
For a clean bulk Nb cavity, this assumption is reasonable ($\xi_0=39$~nm and $\lambda_L=32$~nm~\cite{halbritter4}).
However, for the sputtered film cavity, the literature value may not be accurate because the material has a very fine structure of grain boundaries and many dislocations.
In this study, we aimed to avoid using any literature values and just used the experimental data and BCS theory.
Apparently, RF surface resistance and penetration depth measurement were not sufficient for this purpose.

\section{Magnetometry}\label{sec:mag}
One of the promising complements to the RF measurement of the cavity is the measurement of the upper critical field $B_{\rm c2}$.
This can be done by a small sample representative of the cavity surface.
Such a sample was prepared using a dummy cavity, whose geometry is identical to the real cavity, as a sample holder for the DC bias sputtering.
The samples thus produced are therefore a good representation of the film on the cavity surface.

\subsection{Sample measurement}
\begin{figure}[ht]
\includegraphics[width=90mm]{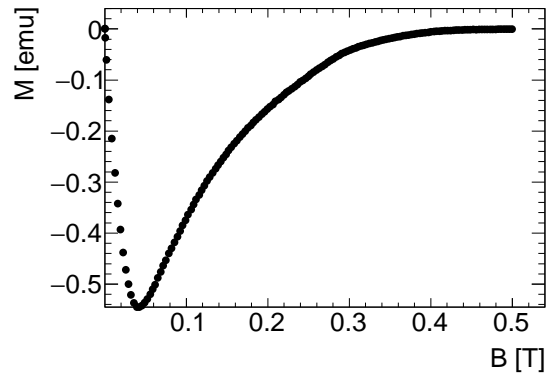}
\caption{
Magnetization curve at $4.5$~K.
\label{fig:M_vs_B}}
\end{figure}

A series of magnetization measurements of one sample was carried out using SQUID-VSM. 
Figure~\ref{fig:M_vs_B} shows a typical result at 4.5~K.
The lower critical field ($B_{\rm c1}$) is strongly affected by the demagnetization factor, and could not be determined precisely by this method. 
The upper critical field $B_{\rm c2}$ is a more robust observable and can be determined relatively precisely as the x-intersect of Fig.~\ref{fig:M_vs_B}.

\subsection{Analysis of upper critical field}
\begin{figure}[ht]
\includegraphics[width=90mm]{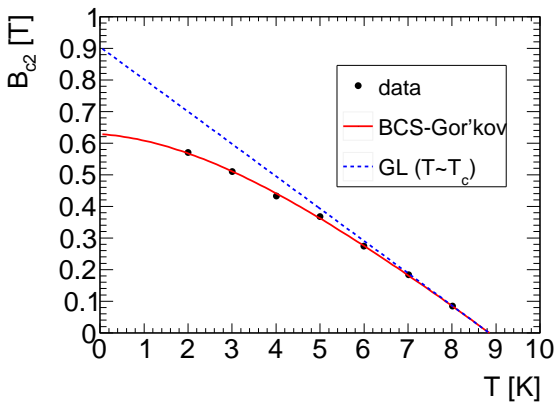}
\caption{
Upper critical field $B_{\rm c2}$ as a function of temperature.
\label{fig:Bc2_vs_T}}
\end{figure}

The temperature dependence of $B_{\rm c2}$ was measured as shown in Fig.~\ref{fig:Bc2_vs_T}.
In the dirty limit,
a full numerical calculation developed by using BCS-Gor'kov formalism~\cite{Bc2_vs_T_1} can be well approximated by an empirical formula~\cite{Bc2_vs_T_2}
\begin{eqnarray}
\label{eq:HHW}
B_{\rm c2}(T) &  =  \frac{B_{\rm c2}(0)}{0.693}h(T) & \nonumber \\
h(T) & = \left(1-\frac{T}{T_{\rm c}}\right) & -  C_1\left(1-\frac{T}{T_{\rm c}}\right)^2 \nonumber \\
     & &  - C_2\left(1-\frac{T}{T_{\rm c}}\right)^4, 
\end{eqnarray}
where $C_1=0.153$, $C_2=0.152$ are fixed, and $B_{\rm c2}(0)$ is the only free parameter.
The number $0.693$ changes to $0.72$ in the clean limit; thus, the uncertainty by ignoring the mean free path is less than $5$\%.
The solid line shown in Fig.~\ref{fig:Bc2_vs_T} is the best fit of (\ref{eq:HHW}).

The dashed line in Fig.~\ref{fig:Bc2_vs_T} shows the Ginzburg-Landau theory tangentially fitted at $T\rightarrow T_{\rm c}$
\begin{eqnarray}
\label{eq:Bc2_GL}
B_{\rm c2}(T) = \frac{\Phi_0}{2\pi\xi_{\rm GL}(T)^2},
\end{eqnarray}
where $\Phi_0$ is the flux quantum, and $\xi_{\rm GL}$ is the Ginzburg-Landau coherence length
\begin{equation}
\label{eq:coherenceGL}
\xi_{\rm GL}(T) \propto \left(1 - \frac{T}{T_{\rm c}} \right)^{-1/2}.
\end{equation}
Note that (\ref{eq:Bc2_GL}) and (\ref{eq:coherenceGL}) are only valid near $T_{\rm c}$.
Therefore, the dashed line over-estimates $B_{\rm c2}$ at $T=0$ by about $30$\% as shown in Fig.~\ref{fig:Bc2_vs_T}.

\subsection{Constraint to the BCS input parameters}
Since expansion of BCS-Gor'kov theory near $T_{\rm c}$ reproduces the Ginzburg-Landau theory~\cite{Gorkov_BCS_to_GL},
the relation between BCS material parameters $(\xi_0, l)$ and $\xi_{\rm GL}(T)$ is obtained by this expansion for arbitrary impurity~\cite{Orland_BCS_to_GL}
\begin{eqnarray}
\label{eq:BCS-GL}
\xi_{\rm GL}(T)&  =  K (\xi_0, l)\frac{\sqrt{R(l)}}{\sqrt{1-T/T_{\rm c}}} \nonumber \\
K(\xi_0, l) & \equiv 0.739 \left[ \frac{1}{\xi_0^2} + 0.882\frac{1}{\xi_0 l}\right]^{-1/2},
\end{eqnarray}
where $1=R(0)<R(l)<R(\infty)=1.17$.
Once the material parameters $(\xi_0, l)$ are provided, (\ref{eq:BCS-GL}), (\ref{eq:Bc2_GL}) and (\ref{eq:HHW}) lead to a theoretical estimation of $B_{\rm c2}(0)$.
One must not forget the factor $\sim 0.7$ to translate Ginzburg-Landau theory near $T_{\rm c}$ to BCS theory at $T=0$.

These formulae can be used as another constraint to the BCS parameters.
Using (\ref{eq:Bc2_GL}) and (\ref{eq:BCS-GL}), we obtain
\begin{equation}
\lim_{t\rightarrow 1} \left(-\frac{dB_{\rm c2}}{dt} \right) = \frac{\Phi_0}{2\pi}K^{-2}\frac{1}{R(l)},
\end{equation}
where $t=T/T_{\rm c}$.
The slope was given by the linear fitting by the dashed line in Fig.~\ref{fig:Bc2_vs_T},
and provided a relation between possible $\xi_0$ and $l$ values.

This constraint between two BCS parameters $\xi_0$ and $l$ is complementary to the surface impedance measurement.
The solid line on Fig.~\ref{fig:xi0_vs_l_combined} shows this additional constraint, 
and the star is the minimum $\chi^2$ satisfying the condition given by the magnetometry.
This method got rid of the last correlation among the parameters and the material parameters were uniquely determined based on experimental data and BCS theory.

\section{Result}
The fitted material parameters are summarized in Table~\ref{tab:fitted_result}.
The first four rows show the result fitted by the BCS theory of surface resistance, penetration depth, and upper critical field.
The next three rows show the parameters determined in advance using different methods.
The last row shows the estimated effective penetration depth.
Cavity 1 had the best $Q_0$ in the series production at the time of writing this report, and cavity 2 performed the worst.

The table also contains three different sample measurements.
Point Contact Tunneling (PCT) measurement~\cite{Tobias1} showed a low superconducting gap $\Delta_0$ while the standard deviation was large. 
This PCT was a local measurement and only probes the surface of the samples.
On the other hand, the coupling determined by the RF measurement was averaged over the cavity surface and penetration depth.

Muon spin rotation $\mu$SR~\cite{Tobias1} resulted in a consistent effective penetration depth calculated by~\cite{halbritter5}
\begin{equation}
\lambda_{\rm eff} = \lambda_{\rm L} \left(1 + \frac{\pi\xi_0}{2l} \right)
\end{equation}
when using $(\xi_0, \lambda_{\rm L}, l)$ obtained by the the BCS data fitting.

The conventional DC 4-contact measurement for Residual-Resistivity Ratio ($\rho({\rm 300K})$ $/\rho({\rm 10K})$ by DC resistance $\rho$) was also done~\cite{alban:sputtering}.
Since the film was deposited on a Cu substrate, a precise 4-contact measurement was very difficult.
Instead, this measurement was done on another film deposited on a quartz sample installed in the same sample holder and simultaneously sputtered with the other samples.
The possible different crystal structure of the films on Cu and quartz gives rise to a doubt on the reliability of this measurement.
However, its result was consistent with BCS fitting ($l\sim2.7\times \rho({\rm 300K})/\rho({\rm 10K})$~\cite{mfp_RRR}).

Reference~\cite{NbCu} is the result of a previous study on 1.5~GHz elliptical cavities coated by DC magnetron sputtering.
As the coating method and cavity geometry are totally different from this study, the different result is not surprising.
The BCS fitting procedure was also different.
In their study, the Ginzburg-Landau parameter of clean limit $\kappa=0.96\lambda_{\rm L}/\xi_{0}$ was fixed at a literature value (0.78),
and surface resistance and penetration depth were also independently fitted.
The fitted parameters were consistent within two standard deviations.

Reference~\cite{halbritter4} is a literature value of bulk Nb in the clean limit.
Previous studies on bulk Nb~\cite{Cornell_PhD} fitted the data with free fitting parameters $(l, \Delta_0/k_{\rm B}T_{\rm c}, T_{\rm c}, R_{\rm res})$,
but fixed $(\xi_0, \lambda_{\rm L})$ at more or less similar values as this column.

The fitting results of cavity 1 are shown on Fig.~\ref{fig:Rs_vs_T} and Fig.~\ref{fig:lambda_vs_T} as solid lines.
They fitted the data very well by the parameters shown in table.~\ref{tab:fitted_result}.

\begin{table*}[t]
  \centering
  \begin{tabular}{lccccccc} \hline
                                  & cavity 1 & cavity 2 & PCT      & $\mu$SR & DC 4-contact & \cite{NbCu} & \cite{halbritter4}\\ \hline
    $\xi_0$~nm                    & 29(7)    &  28(7)   &          &         &              & 36(4)       & 39   \\
    $\lambda_{\rm L}$~nm          & 26(7)    &  28(7)   &          &         &              & 29(3)       & 32   \\
    $l$~nm                        & 99(25)   &  139(30) &          &         & 95(27)       & 5-1000      &      \\
    $\Delta_0/k_{\rm B}T_{\rm c}$ & 1.7(1)   &  1.5(1)  & 1.6(6)   &         &              & 1.87        & 1.75-1.93 \\ \hline
    $R_{\rm res}$~n$\Omega$       & 11.8     &  9.0     &          &         &              &             &      \\
    $T_{\rm c}$~K                 & 9.6      &  9.6     &          &         &              & 9.54        & 8.95-9.2      \\
    $T_0$~K                       & 6.3      &  7.1     &          &         &              &             &      \\ \hline
    $\lambda_{\rm eff}$~nm        & 31       &  32      &        & 29(5)   &              &        &      \\
  \end{tabular}
  \caption{Material parameters determined by this study compared with references.}
  \label{tab:fitted_result}
\end{table*}

\section{Discussion}
The fitting result showed slightly shorter coherence length in sputtered Nb film than bulk Nb.
The reason is not clear, but we excluded that this comes from the smaller grain size in the film than bulk.
This is because picture analysis of the crystal structure showed that the averaged grain size near the surface is of the order of 100~nm, and still much longer than the fitted coherence length.
Instead, the observed grain size is comparable to the fitted mean free path.
The crystal grains in our cavity appear to be scattering centers of normal conducting electrons.

The fitted $\Delta_0/k_{\rm B}T_{\rm c}$ was weaker than the literature value, and was also weaker in cavity 2 than cavity 1.
The PCT measurement~\cite{Tobias1} showed a rather wide spread in $\Delta_0$ for the DC-bias sputtered samples representative of HIE-ISOLDE cavities, compared with other sputtering techniques.
There were weak superconducting junctions or even non-superconducting junctions over the surface.
In cavity 2,  after chemical processes, several cracks were found in the heat affected zone of the welding on the substrate.
The cavity was coated without any particular treatment on these cracks.
This might indicate that a contamination caused by the chemical polish was left in the cracks and eventually resurfaced during the sputtering process
when the substrate was heated up to $620^{\circ}$C.
Also, the film deposited on the crack might grow inhomogeneously and could result in a lower superconducting gap.

The cavities and analysis presented in Ref.~\cite{NbCu} were totally different and not easy to compare with this study.
As this study showed, magnetometry provides complementary information to the RF measurement.
There was a study by another team about $B_{\rm c2}$~\cite{Bc2_DCM_1} cited by a couple of different works~\cite{Bc2_DCM_2}\cite{Padamsee}.
They have sometimes measured $B_{\rm c2}$ to be higher than $3$~T.
If their result was correct, and BCS theory is still applicable, this means very short $\xi_0$ and $l$ without affecting $T_{\rm c}$.
For this reason too, a dedicated and more systematic study comparing different coating methods is of interest.

\section{Conclusion}
The material parameters of the Nb sputtered cavity were determined only by the experiment and BCS theory without any literature values.
Strong correlations among the parameter estimators were pointed out, and were eliminated by using surface resistance, penetration depth, and the upper critical fields.
Some of the fitted parameters showed difference from the literature of bulk Nb in the clean limit.
The method shown in this paper is general and can be a standard procedure for the performance analysis of superconducting cavities.

\section*{Acknowledgement}
We gratefully acknowledge the contribution of our colleagues A.~Sublet, S.~Teixeira, and M.~Therasse for their support in cavity preparation and testing. 
M.~Eisterer and T.~Sch\"afer carried out the magnetic measurements with SQUID at Technical University of Vienna. 
Our special thanks go to N. Shipman for useful discussion.
We warmly thank all the technical staff at CERN for their invaluable help.

\appendix
\section{RF measurement}\label{sec:RF_meas}
The cavity performance was fully evaluated by RF measurements without using any calorimetric methods.
During the measurement, the cavity should be well locked at on-resonance.
First, the quality factor of the pick-up port $Q_{\rm pick}$, a geometrical constant during the measurement, was calibrated as follows.
The time constant of the energy decay $\tau$ was evaluated at low field (typically $E_{\rm acc}<1.0$~MV/m) where the non-linear phenomenon distorting exponential decay can be neglected (uncertainty was less than 5\%).
The loaded quality factor $Q_{\rm L}$ was then directly obtained from $\tau$ as
\begin{equation}
Q_{\rm L} = \omega\tau,
\end{equation}
where $\omega$ is the angular resonant frequency $\omega = 2 \pi f$.
With the same configuration as the field-decay measurement, the steady state powers $\left(P_{\rm f}, P_{\rm r} , P_{\rm t}\right)$ were measured, 
and the coupling coefficient was calculated by
\begin{equation}
\beta = \frac{1\pm\sqrt{P_{\rm r}/P_{\rm f}}}{1\mp\sqrt{P_{\rm r}/P_{\rm f}}},
\end{equation}
where the upper sign is used for over-coupling, and the lower sign is used for under-coupling case.
The power consumption in the cavity $P_{\rm c}$ is
\begin{equation}
P_{\rm c} = P_{\rm f} - P_{\rm r} - P_{\rm t}.
\end{equation}
The coupling coefficient of the pick-up port $\beta_{\rm pick}$ was also evaluated as
\begin{equation}
\beta_{\rm pick} = \frac{P_{\rm t}}{P_{\rm c}}.
\end{equation}
Then, the cavity quality factor  was calculated by
\begin{equation}
Q_{0} = Q_{\rm L} \left( 1+\beta+\beta_{\rm pick} \right),
\end{equation}
and the quality factor of the pick-up port is given by
\begin{equation}
Q_{\rm pick} = \frac{Q_0P_{\rm c}}{P_{\rm t}}.
\end{equation}
Three measurements at over-coupling, critical-coupling, and under-coupling were done, and they resulted in consistent $Q_{\rm pick}$ within 10\% uncertainty.

Once $Q_{\rm pick}$ was determined, $Q_0$ and $E_{\rm acc}$ at steady states of arbitrary $\left(P_{\rm f}, P_{\rm r} , P_{\rm t}\right)$ were easily obtained.
From the transmitted power at the pick-up port
\begin{equation}
U = \frac{Q_{\rm pick}P_{\rm t}}{\omega}.
\end{equation}
The cavity quality factor is 
\begin{equation}
Q_{0} = \frac{\omega U}{P_{\rm c}}.
\end{equation}
The cavity field can be calculated by 
\begin{equation}
E_{\rm acc} = \sqrt{U/\kappa},
\end{equation}
where $\kappa$ was evaluated by the RF simulation, and is $0.207\, {\rm J(MV)^{-2}m^{2}}$.
For accurate measurement, the fundamental power coupler was always moved to the near critical coupling position ($P_{\rm r}/P_{\rm f}\ll 5$\%)
so that the standing-wave in the RF power cable is minimized.

The method explained here is accurate if 
the directional coupler splitting forward and reflected power has good isolation.
The main source of systematic uncertainty is calibration of the cable attenuation, especially of the forward sampling line
because other powers are practically zero near the critical coupling condition.
The accuracy is typically no better than $10$\% for $Q_0$ and $5$\% for $E_{\rm acc}$ in each calibration.
This is an absolute systematic error commonly shared by all the data points of the same calibration.
There are other types of errors, such as human error, fluctuation of the power,  and phase error, but can be typically one order of magnitude smaller.



\theendnotes

\end{document}